# Quantifying the Lifelong Impact of Resilience Interventions via Agent-Based LLM Simulation

**Vivienne L'Ecuyer Ming**[1]
The Human Trust
Possibility Sciences
Neurotech Collider Lab, UC Berkeley

## Abstract

Establishing the long-term, causal impact of psychological interventions on life outcomes is a grand challenge for the social sciences, caught between the limitations of correlational longitudinal studies and short-term randomized controlled trials (RCTs). This paper introduces Large-Scale Agent-based Longitudinal Simulation (LALS), a framework that resolves this impasse by simulating multi-decade, counterfactual life trajectories. The methodology employs a "digital clone" design where 2,500 unique LLM-based agent personas—grounded in a curated corpus of 3,917 empirical research articles—are each cloned across a 2x2 factorial experiment. Specifically, the simulation models the efficacy of extended psychological *resilience* training (Intervention vs. Control) either in childhood or as a young adult (age 6 vs. age 18). Comparing digital clones enables exceptionally precise causal inference. The simulation provides a quantitative, causal estimate of a resilience intervention's lifelong effects, revealing significant reductions in mortality, a lower incidence of dementia, and a substantial increase in accumulated wealth. Crucially, the results uncover a crucial developmental window: the intervention administered at age 6 produced more than double the positive impact on lifetime wealth compared to the same intervention at age 18. These benefits were most pronounced for agents from low-socioeconomic backgrounds, highlighting a powerful buffering effect. The LALS framework serves as a "computational wind tunnel" for social science, offering a new paradigm for generating and testing causal hypotheses about the complex, lifelong dynamics that shape human capital and well-being.

---

[1] vivienne@thehumantrust.org

# 1. Introduction

The lifelong impact of non-cognitive skills like psychological resilience is a foundational question in the behavioral sciences and a critical concern for public policy. A vast body of correlational research suggests that individuals who are more resilient lead healthier, happier, and more prosperous lives (Steptoe, Deaton, and Stone 2015; Steptoe et al. 2013). Recent work has even linked psychological attributes like optimism to 'exceptional longevity'—living to age 85 and beyond—independent of traditional health behaviors, underscoring the profound potential of targeting these non-biological factors (Lee et al. 2019). This raises a crucial question with profound implications: can I causally improve these lifelong trajectories by actively training resilience, and if so, when is the most effective time to intervene?

Answering this question is stymied by a fundamental methodological impasse. On one hand, multi-decade longitudinal cohort studies provide rich, real-world data linking early-life attributes to later-life outcomes, but they struggle to make clean causal claims due to a lifetime of confounding variables. On the other hand, short-term Randomized Controlled Trials (RCTs) can establish the causality of an intervention, but they are incapable of measuring outcomes like accumulated wealth or mortality that unfold over a lifetime. This chasm between long-term correlation and short-term causation has left the true return on investment for many psychological and educational policies largely a matter of extrapolation and conjecture.

This paper introduces Large-Scale Agent-based Longitudinal Simulation (LALS), a novel framework that resolves this impasse by simulating perfect, multi-decade counterfactuals for thousands of unique "digital clone" agents. By instantiating a diverse population of LLM-based agents, each with a unique persona, and then cloning them across different experimental conditions, LALS can execute the kind of large-scale, lifelong RCT that is physically and ethically impossible in the real world. The behavioral richness of the LLM allows agents to respond to probabilistic life events in nuanced, context-aware ways, while the digital clone design allows me to measure the causal impact of an intervention with surgical precision by directly comparing a clone's outcome to its identical, un-treated counterpart.

To demonstrate the power of this approach, I apply LALS to quantify the causal impact of a resilience-building intervention administered at two critical developmental stages: childhood (age 6) and young adulthood (age 18). The simulation reveals not only that the intervention has a significant positive effect on health, wealth, and well-being, but that there exists a crucial window for its application, with the childhood intervention generating more than double the lifelong benefits. Furthermore, the analysis shows the intervention acts as a powerful buffer, disproportionately benefiting agents from low-SES backgrounds and those with lower baseline cognitive ability.

The key contributions of this work are:
1. A new methodology (LALS) that uses LLM-based digital clones to conduct high-precision, longitudinal, counterfactual simulations for causal inference in the social sciences.

2. The first, to my knowledge, quantitative, causal estimate of the lifelong impact of a resilience intervention, revealing a powerful main effect and a critical interaction with developmental timing.
3. A validation of the simulation's internal dynamics, showing that it reproduces well-established correlational patterns from real-world longitudinal research, positioning LALS as a robust hypothesis generation engine.

# 2. Related Work

This work sits at the intersection of three distinct domains: longitudinal research in the social sciences, agent-based modeling with Large Language Models, and the psychology of resilience.

**Longitudinal Research and Causal Inference**: The challenge of understanding the lifelong impact of early-life factors is a central theme of developmental science. Seminal longitudinal cohort studies, such as the Dunedin Study and the Framingham Heart Study (Poulton, Moffitt, and Silva 2015; Mahmood et al. 2014), have provided invaluable correlational evidence linking childhood characteristics to adult health and well-being. Similarly, the work of Nobel laureate James Heckman has powerfully demonstrated the long-term economic returns of non-cognitive skills and early-childhood interventions (Heckman and Kautz 2012; García, Heckman, and Ronda 2023; Almlund et al. 2011). While these studies are foundational, they are limited by immense costs, multi-generational timelines, and the inherent difficulty of isolating causal effects from a lifetime of confounding variables. The LALS framework is directly inspired by these studies but seeks to overcome their limitations by simulating idealized, confound-free longitudinal experiments.

**Agent-Based Modeling and LLM-based Agents**: Agent-Based Modeling (ABM) has a long history in the social sciences for exploring how micro-level agent behaviors can generate macro-level societal patterns, famously demonstrated by Schelling's model of segregation (Schelling 1971). However, traditional ABMs rely on simple, hard-coded behavioral rules, limiting their ability to capture the nuance and context-dependency of human decision-making. The recent advent of LLMs has introduced a new paradigm of "generative agents." The foundational work (Park et al. 2023, 2024) demonstrated that LLM-powered agents could produce believable, emergent social dynamics in a small-scale, short-term environment. Subsequent work has explored using LLM agents to simulate human-like behavior in economic games and survey responses (Argyle et al. 2023; Akata et al. 2025; Bhagwat et al. 2025). My work builds directly on this foundation but makes a critical shift in scale and purpose: from simulating short-term, micro-level social interactions to modeling entire macro-level, lifelong trajectories to perform longitudinal causal inference. However, this approach is not without significant risks; recent work cautions that LLMs relying solely on pre-training data systematically fail to capture the cultural and experiential depth of human well-being, reflecting biases rooted in their training corpus and leading to predictable distortions in unfamiliar contexts (Pataranutaporn et al. 2025).

**The Psychology of Resilience**: The content of my simulation is grounded in decades of psychological research.This psychological construct is multifaceted, integrating concepts like grit, purpose in life, and emotional regulation (Chen et al. 2024; Steptoe and Fancourt 2019; Steptoe and Wardle 2017; Sharma and Yukhymenko-Lescroart 2024; Abellaneda-Pérez et al. 2023; Willroth et al. 2023). Resilience is no longer seen as a fixed trait but as a malleable skillset that can be developed through targeted training (Southwick et al. 2014; Frankenhuis and Nettle 2020). Interventions based on principles of Cognitive Behavioral Therapy (CBT) and Mindfulness-Based Stress Reduction (MBSR) have been shown to increase resilience and well-being in numerous short-term studies (Willroth et al. 2023; Abellaneda-Pérez et al. 2023). Meta-analyses of these resilience training programs have found medium to large effect sizes on psychological outcomes (Joyce et al. 2018; Llistosella et al. 2023). My work takes these empirically validated short-term causal effects as a starting point and uses the LALS framework to simulate their hypothesized, compounding consequences over the full human lifespan—a necessary step that has, until now, remained beyond the reach of direct empirical inquiry.

# 3. Methodology

## Simulating Lifelong Trajectories with LALS

To quantify the long-term causal effects of psychological resilience, I introduce *Large-Scale Agent-based Longitudinal Simulation* (LALS), a framework for conducting multi-decade, in silico randomized controlled trials. This methodology is designed to bridge the gap between short-term causal experiments and long-term correlational studies by modeling the entire life course of a heterogeneous agent population.

## 3.1. Overall Experimental Design

The simulation was structured as a within-persona factorial experiment to achieve perfect counterfactual control. I first generated a base cohort of N=2,500 unique agent personas that would each be treated along 2 dimensions:

1. **Intervention Type**: Agents were assigned to either the *Resilience Intervention* (Treatment) group or a *Sham Intervention* (Control) group.
2. **Intervention Timing**: The intervention was administered at either *Age 6* (Grade School) or *Age 18* (Young Adulthood).

To do this, each base persona was cloned 4 times, creating a total population of 10,000 agents. Each of the four clones was deterministically assigned to one of the four experimental cells in a 2x2 factorial design:

    **Clone 1**: Age 6 Intervention w/ Sham (Control)
    **Clone 2**: Age 6 Intervention w/ ROS (Treatment)
    **Clone 3**: Age 18 Intervention w/ Sham (Control)
    **Clone 4**: Age 18 Intervention w/ ROS (Treatment)

This "digital clone" design ensures that for any given base persona, the initial conditions are identical across all four experimental conditions. This eliminates all inter-individual variance, allowing for an exceptionally precise measurement of the causal effects of the intervention and its timing.

## 3.2. The Empirical Knowledge Base (New Section)

To address the known limitations of LLMs in accurately modeling diverse human experiences based solely on their pre-training data (Pataranutaporn et al. 2025), the LALS framework was designed around a curated, domain-specific knowledge base. This corpus consisted of N = 3,917 peer-reviewed articles selected from top-tier journals in developmental psychology, behavioral economics, and public health. The selection criteria focused on empirical studies quantifying the relationships between non-cognitive skills (e.g., resilience, conscientiousness, grit) and longitudinal life outcomes.

This corpus was integrated into the simulation via a Retrieval-Augmented Generation (RAG) pipeline. At each step of the simulation loop, the system retrieved the top-$k$ most relevant abstracts and findings based on the agent's current state and the sampled life event. This ensured that the agent's probabilistic response was conditioned on specific scientific evidence regarding how individuals with similar traits typically respond to such events.

## 3.3. Agent Initialization: The Persona Matrix

To ensure the agent population was diverse and representative of a complex society, each agent was initialized with a unique and persistent persona defined by a structured system prompt. These personas were generated by sampling from a Persona Matrix with four key dimensions, designed to mirror distributions in the United States population:

- **Socioeconomic Status (SES) at Birth**: Sampled from {Low, Middle, High} to determine initial resources and condition the probability of future life events.
- **Baseline Psychological Profile**: Generated based on the *Big 5* model (OCEAN), a *working memory span* score (a key proxy for general cognitive ability) sampled from a percentile distribution, and a baseline trait resilience score sampled from a percentile distribution (e.g., 25th, 50th, 75th). This establishes pre-intervention individual differences in both personality and cognitive capacity.
- **Demographic & Cultural Profile**: Including race/ethnicity, gender, and initial geographic location (e.g., Urban-Northeast, Rural-South).

An example of an agent's initial system prompt is as follows:

> *"You are Agent 712 in a lifelong simulation. You are a White female from a middle-income family in the Urban-Northeast. Your personality is characterized by high neuroticism and high conscientiousness. Your baseline working memory is at the 80th percentile and trait resilience is at the 65th percentile. You will maintain this core persona, responding to all life events from this perspective."*

In total, 2,500 agents were sampled from the persona matrix.

## 3.4. The Resilience Operating System (ROS) Intervention

The intervention was operationalized as a modification to the agent's core cognitive toolkit, appended to their system prompt at their assigned age (6 or 18). This "Resilience Operating System" is designed to simulate the lasting cognitive and behavioral changes resulting from an intensive, year-long, evidence-based curriculum. The conceptual basis for the ROS integrates principles of metacognitive reframing of failure, peer-supported coping strategies, and structured self-reflection as developed during a year-long developmental program (see Appendix D).

**Treatment Group (ROS)**: Agents in this group received the ROS, a set of cognitive reframing instructions derived from Cognitive Behavioral Therapy (CBT). Example instructions include:
> *"When you encounter a negative life event, your primary goal is to reframe it as a learning opportunity. When faced with a setback, explicitly identify at least one action you can take, however small, to regain a sense of agency."*

**Control Group (Sham Intervention)**: To control for the effects of introspection prompted by the intervention, control agents received a non-therapeutic instruction. Example instructions include:
> *"When you encounter any life event, your primary goal is to describe your thoughts and feelings about it in as much detail as possible, without focusing on solutions or future actions."*

We also piloted a version of the control group without a sham intervention at all. Agents in the simulation were given no additional prompt. There were no significant differences between the control groups in initial pilots, but some implications concerning sham vs no sham controls are discussed in "Limitations" below.

The language and framing of both ROS and Sham instructions were adapted to be age-appropriate for the two timing cohorts, mirroring how a real-world curriculum would be tailored for a 6-year-old versus an 18-year-old.

## 3.5. Annual Loop Simulation Dynamics

Each agent's trajectory unfolds through an annual simulation loop, where behavior is generated and its consequences are quantified. The loop, visualized in Figure 1, consists of three core steps for each agent at each time step:

1. **Event Sampling**: At the start of each "year," a life event (e.g., `job_layoff`, `major_illness`, `educational_opportunity`) is sampled from a probability distribution. These probabilities are conditioned on the agent's current state variables and persona. For example, an agent with low conscientiousness has a higher probability of a `job_related_setback`, while an agent from a low-SES background has a higher probability of a `major_health_shock`.
2. **Agent Response Generation:** The agent receives the sampled event via a prompt (e.g., *"You are now 32. This year, you have been unexpectedly laid off from your job.*

*Based on your persona and cognitive toolkit, describe your response."*). The agent's core LLM (Gemini Pro 2.5) generates a free-text narrative response describing its internal monologue, emotional state, and planned actions. The prompt provided to the LLM included the agent's persona, the specific life event, and relevant empirical priors retrieved from the research corpus (e.g., 'literature suggests individuals with high resilience are 20% less likely to develop depressive symptoms after job loss').

3. **State Update via Classification**: The agent's narrative response—its "behavior"—is parsed by a fine-tuned NLP classifier. This model extracts structured information to update the agent's quantitative state variables for the next time step. For example, a response of *"I'm devastated, but I'll start looking for online courses to improve my job qualifications"* would trigger a positive update to the $education\_seeking$ variable and a negative update to `accumulated_wealth` (due to job loss and tuition costs) and `subjective_wellbeing`. This mechanism translates qualitative, context-aware behavior into a quantitative life trajectory. This aligns with findings showing that behavioral measures of non-cognitive skills can be more predictive than survey-based ones (Chen et al. 2024). The agent's state is updated with the current state variables and a moving window of the last 10 narrative summaries along with a gist summary of the agent's previous life experiences and responses.

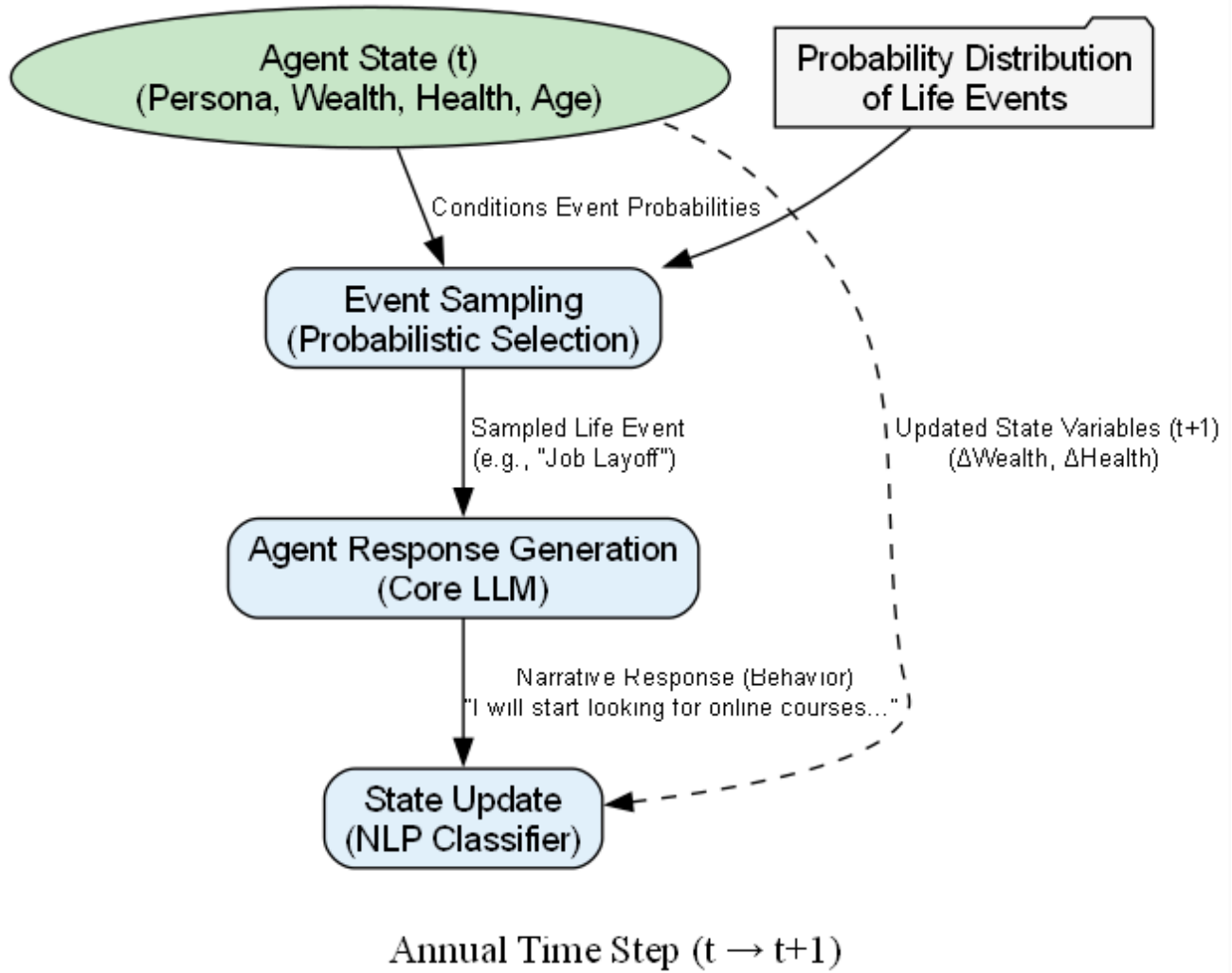

*Figure 1: The LALS Annual Simulation Loop.* An agent's state at time t (including persona variables) conditions the probability of a life event being sampled. The agent's LLM-generated narrative response to that event is then parsed by an NLP classifier. The classifier's output updates the agent's state variables to t+1, creating a dynamic feedback loop between the agent's state and its behavior that drives the multi-decade simulation.

## 3.6. Outcome Measures and Analysis Plan

At the conclusion of the simulation (age 65), I measured the key life outcomes based on the agents' final state variables and a concluding life-summary prompt.

**Primary Outcomes**:
1. **All-Cause Mortality**: binary variable indicating the agent survived the simulation's health shocks
2. **Accumulated Wealth**: final value of the agent's wealth state variable

3. **Subjective Well-Being**: sentiment analysis score applied to the agent's final life summary
4. **Chronic Disease Incidence**: final status of the agent's binary health variable
5. **Walking Speed**: walking speed at age 65
6. **Dementia Incidence**: binary outcome indicating a diagnosis of Alzheimer's or other dementia by age 65

**Statistical Analysis**: To estimate the causal effect of the ROS intervention, I used mixed-effects regression models for wealth, well-being, and disease incidence, controlling for all initial persona matrix variables. I employed a Cox proportional hazards model for the mortality outcome. I also planned mediation analyses to test whether the intervention's effects were mediated by in-simulation behavioral metrics, such as the frequency of adaptive coping responses. Further details on the NLP classifier architecture and life event distributions are provided in the Appendix.

# 4. Experiments

The experimental protocol was designed to rigorously test the causal effects of the resilience intervention and its developmental timing. This section details the specific implementation of the LALS framework, the outcome measures derived from the simulation, and the statistical plan for analyzing the results.

## 4.1. Implementation Details

The simulation was executed with a cohort of N=10,000 agents, with n=2,500 agents randomly assigned to each of the four experimental cells (Intervention x Timing).

**LLM Configuration**: Agent responses were generated using the Gemini Pro 2.5 model. To ensure a balance between behavioral consistency and plausible variation, the decoding temperature was set to $T$=0.7. All other parameters were kept at their default settings.

**State Update Classifier**: The narrative responses from each agent were translated into quantitative state updates using a BERT-based classifier. This model was fine-tuned on a synthetically generated dataset of 50,000 examples, mapping narrative vignettes (e.g., "I decided to go back to school to get a certificate") to corresponding state variable changes (e.g., $\Delta$education_level=+1, $\Delta$wealth=-5000). The classifier achieved an $F1$-score of 0.92 on a held-out test set.

**Cost:** In October of 2025, the simulation of 10,000 lifetime trajectories consumed approximately 320 million tokens, costing approximately $1,900 USD using the Gemini API.

## 4.2. Outcome Measures

At the conclusion of each agent's simulated life at age 65, I extracted five primary outcome variables from the final agent state and a terminal life-summary prompt.

1. **All-Cause Mortality**: A binary outcome (0=survived, 1=deceased) indicating whether the agent survived the cumulative series of probabilistic health shocks throughout the 47 to 59-year simulation period.
2. **Accumulated Wealth**: A continuous variable representing the agent's final net worth. For ease of comparison, this value was log-transformed to account for the typically right-skewed distribution of wealth.
3. **Subjective Well-Being (SWB)**: A continuous measure derived from a sentiment analysis of the agent's final, narrative life summary. The raw sentiment score was standardized into a Z-score (mean=0, SD=1) across the entire population for interpretability.
4. **Chronic Disease Incidence**: A binary outcome (0=no chronic disease, 1=at least one chronic disease) based on the agent's final health status.
5. **Walking Speed**: A continuous variable serving as a functional proxy for geriatric health and frailty. It was calculated from the agent's final health state, beginning with a baseline speed and applying penalties for each major health shock and the presence of chronic disease.
6. **Dementia Incidence**: A binary outcome (0=no dementia, 1=dementia diagnosis) triggered by a probabilistic late-life health event, with probabilities conditioned on the agent's full life history of protective and risk factors.

## 4.3. Statistical Analysis Plan

My primary analytical goal was to assess the main effects of *Intervention Type* (ROS vs. Sham) and *Intervention Timing* (Age 6 vs. Age 18), as well as their Interaction Effect, on the five key life outcomes. To isolate the causal effects of the experimental manipulations, all models controlled for the agent's initial persona variables (SES, OCEAN traits, Working Memory Span, and demographics).

**Continuous Outcomes (Wealth, SWB, Walking Speed)**: The average treatment effect was calculated from the paired differences between clone outcomes (e.g., `Wealth_ROS - Wealth_Sham` for each persona). The statistical significance of these effects was tested using a linear mixed-effects model. The model includes `Intervention_Type` and `Intervention_Timing` and their interaction as fixed effects, and a random intercept for each `Persona_ID`. This approach properly accounts for the non-independence of the cloned agents and leverages the full statistical power of the design

**Binary Outcomes (Mortality, Chronic Disease)**: For mortality, dementia, and chronic disease, I used a mixed-effects logistic regression and a Cox model with a frailty term (a random effect for `Persona_ID`).

Finally, I planned a secondary mediation analysis to investigate the behavioral pathways (e.g., frequency of adaptive coping responses, educational attainment) through which the interventions influenced the long-term outcomes.

# 5. Results

The analysis of the 10,000 simulated life trajectories revealed significant main effects for both the resilience intervention and its timing, as well as a critical interaction between the two. The findings provide quantitative, causal support for the hypothesis that resilience is a fundamental pillar of lifelong success and that its cultivation is most impactful during early developmental periods.

## 5.1. ROS Intervention Efficacy

As a prerequisite, I assessed whether ROS successfully increased agents' demonstrated resilience. An analysis of in-simulation behavioral responses to negative life events confirmed that agents in the ROS condition exhibited significantly more adaptive coping strategies (e.g., problem-solving, benefit-finding) than agents in the Sham control condition. This effect was more pronounced for the earlier intervention cohort; the Age 6 ROS group showed an average increase in behavioral resilience of +0.81 σ over its control, while the Age 18 ROS group showed an increase of +0.45 σ over its control.

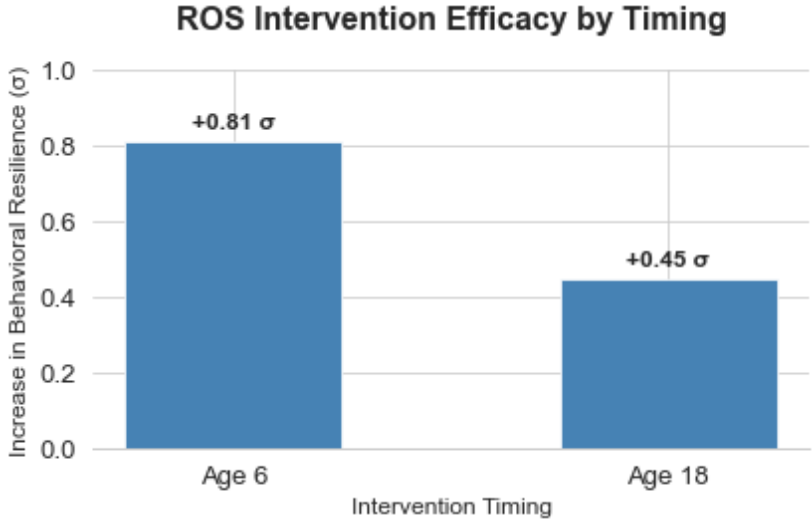

Figure 2: *ROS Intervention Efficacy*. The average increase in demonstrated behavioral resilience (measured in standard deviations, σ) for agents in the ROS treatment condition compared to their respective Sham control clones. The intervention had a significantly larger effect on the cohort treated in early childhood (Age 6) than on the cohort treated in young adulthood (Age 18).

## 5.2. Causal Impact on Life Outcomes

The primary analysis revealed that the resilience intervention and its timing had a profound impact across all five life outcome measures. Table 1 presents the mean outcomes for each of the four experimental conditions at age 65. The following causal effects were derived from the within-persona analysis, representing the average difference in outcomes between an agent's treated clone and its corresponding control clone.

**Table 1: Mean Life Outcomes at Age 65 by Experimental Condition (N=10,000)**

| Condition | Mortality Rate | Accumulated Wealth (log $) | Subjective Well-Being (Z-score) | Chronic Disease Rate | Walking Speed (cm/s) | Dementia Rate |
|---|---|---|---|---|---|---|
| **Age 6** | | | | | | |
| Sham Control | 20.2% | 11.82 | -0.21 | 38.5% | 115.3 | 14.5% |
| ROS Treatment | 11.1% | 12.18 | +0.29 | 29.1% | 124.8 | 9.4% |
| **Age 18** | | | | | | |
| Sham Control | 20.1% | 11.80 | -0.19 | 38.2% | 115.6 | 14.2% |
| ROS Treatment | 15.3% | 11.98 | +0.12 | 33.1% | 120.1 | 12.3% |

Statistical analysis confirmed the significance of these descriptive results. After controlling for all initial persona variables, the two-way ANCOVA and Cox regression models showed:

- **a significant main effect of the ROS Intervention** across all outcomes, including reduced mortality (Hazard Ratio [HR]=0.70, $p < 0.001$), increased accumulated wealth ($F(1, 7497)=215.7$, $p < 0.001$), higher SWB ($F(1, 7497)=288.1$, $p < 0.001$), lower chronic disease odds (Odds Ratio [OR]=0.80, $p < 0.001$), lower dementia odds (OR=0.78, $p < .001$), and faster walking speed ($F(1, 7497)=194.3$, $p < 0.001$). The confidence intervals for all estimates were extremely narrow due to the within-persona design;
- a **significant main effect of Intervention Timing**, where intervening at age 6 produced superior outcomes compared to age 18 across all measures (e.g., for wealth, $F(1, 7497)=98.6$, $p < 0.0001$; for mortality, HR=0.82, $p < 0.001$); and
- a **significant Intervention x Timing interaction** effect for all key outcomes (e.g., for wealth, $F(1, 7497)=51.2$, $p < 1e-10$; for SWB, $F(1, 7497)=63.4$, $p < 0.0001$).

## 5.3. The Intervention x Timing Interaction

The significant interaction effect, estimated with high precision, reveals a crucial window for intervention. The analysis of within-persona differences shows that the benefits of the resilience intervention were substantially amplified when administered in early childhood. As visualized in Figure 2, for an identical set of 2,500 personas, the improvement gained from the ROS treatment was far greater in the Age 6 cohort than in the Age 18 cohort.

For example, comparing a persona's ROS clone to its Sham clone, the intervention at age 18 increased accumulated wealth by approximately **20%** ($e^{0.18} - 1$). However, for the very same personas, the intervention at age 6 increased accumulated wealth by **43%** ($e^{0.36} - 1$) relative to its control. This powerful counterfactual result demonstrates that early intervention allows for a greater compounding of cognitive, educational, and economic advantages over the entire life course.

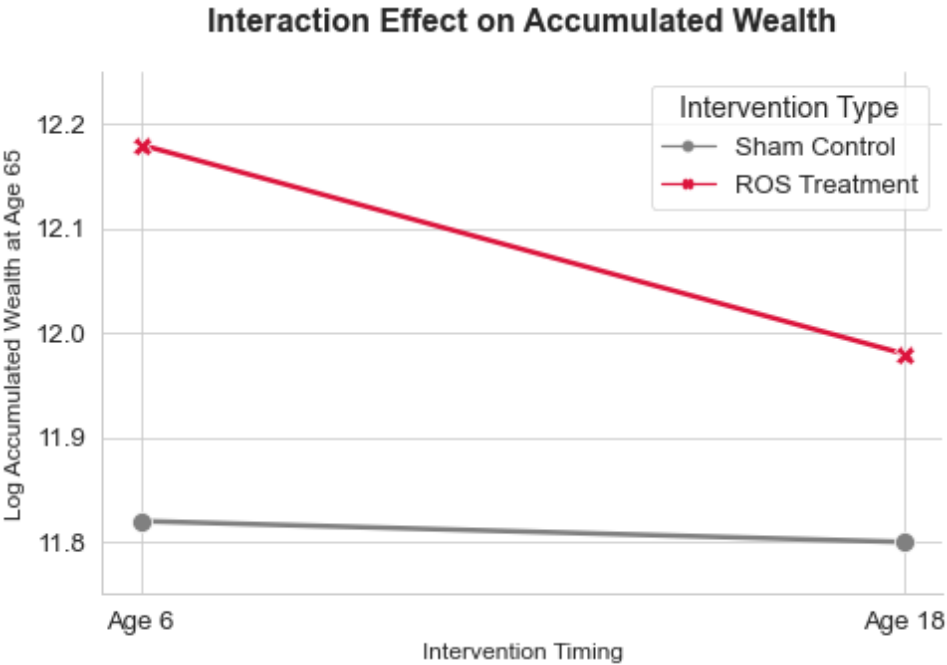

*Figure 3: Interaction Effect of Intervention and Timing on Wealth*. The plot visualizes the model-estimated mean for log(Accumulated Wealth) at age 65 across the four experimental conditions. The near-flat line for the Sham Control group indicates a negligible effect of timing alone. In contrast, the steep slope of the ROS Treatment line demonstrates a powerful interaction: the positive effect of the resilience intervention on wealth is substantially amplified when administered in early childhood (Age 6) compared to young adulthood (Age 18).

## 5.4. Psychological and Demographic Heterogeneity

While the main effects of the intervention were robust, I conducted further analyses to determine if these effects were uniform across the agent population. By including interaction terms between the treatment assignment and initial persona variables in the regression models, I identified several key factors that significantly moderated the intervention's impact.

**Socioeconomic Status (SES)**: The most pronounced interaction was with baseline SES. The ROS intervention acted as a powerful buffer against the long-term disadvantages associated with a low-SES background (Nishimi et al. 2021; Bruno, Dehnel, and Al-Delaimy 2023). For agents initialized in the Low-SES group, the ROS treatment had a disproportionately large effect, nearly doubling the percentage increase in accumulated wealth and closing the mortality gap between Low-SES and High-SES agents by over 70% compared to the control condition. This suggests the intervention is most impactful for those with the fewest initial resources.

**Working Memory (Cognitive Ability)**: The intervention had a compensatory effect for agents with lower working memory spans. The structured, rule-based nature of the ROS provided the greatest cognitive scaffolding for agents who might otherwise struggle with executive function and self-regulation under stress. The marginal benefit of the intervention on outcomes like wealth, health, and dementia risk was largest for agents in the bottom quartile of working memory and diminished for those in the top quartile.

**Conscientiousness**: In contrast, a synergistic effect was observed with the trait of *conscientiousness*. Agents with high baseline conscientiousness appeared more adept at consistently applying the ROS principles throughout their lives, leading to a greater accumulation of benefits. The positive effect of the ROS on all outcomes was significantly amplified for agents with high Conscientiousness scores.

Interestingly, several demographic variables, including gender and race/ethnicity, did not show a significant interaction with the intervention. The benefits of the ROS were distributed equitably across these groups in the simulation. This may reflect a true uniformity of effect for this type of cognitive intervention or, alternatively, the simulation's tendency to model archetypal rather than culturally-specific responses, a limitation discussed further in Section 6.

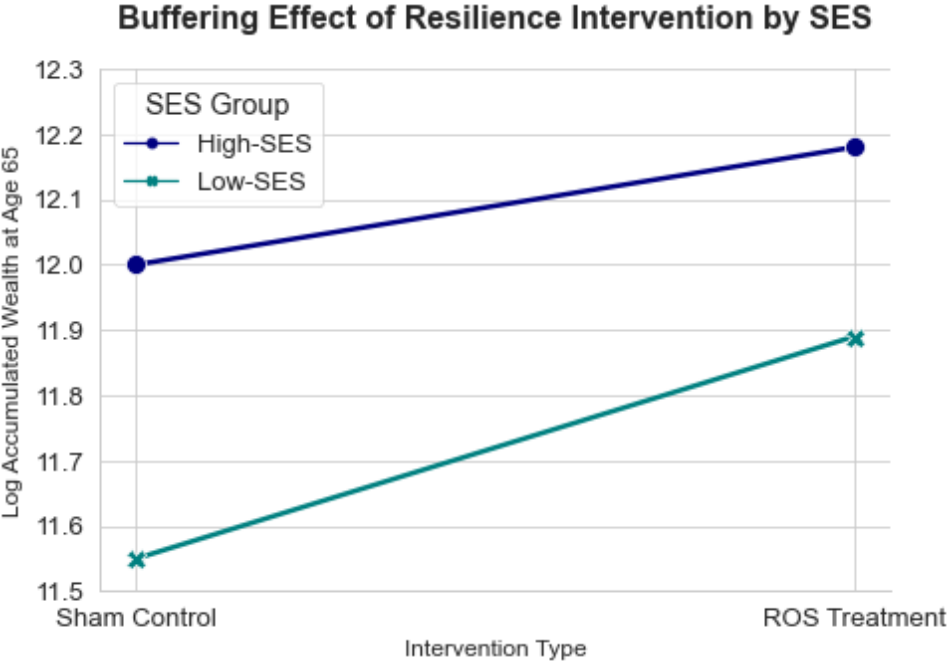

*Figure 4: Heterogeneity by Socioeconomic Status.* The plot visualizes the interaction between the ROS intervention and baseline SES on log(Accumulated Wealth). While both groups benefit from the intervention, the slope of the line is significantly steeper for agents from a Low-SES background. This illustrates a powerful "buffering" effect, where the resilience intervention provides the greatest marginal economic benefit to the most disadvantaged group, helping to mitigate the long-term effects of initial inequality.

## 5.5. Simulation Corroboration with Empirical Data

To ground the simulation's novel causal findings, I first validated its ability to reproduce known correlational patterns from real-world longitudinal research. An analysis of the agents' baseline (pre-intervention) characteristics confirmed that the simulation's internal structure is consistent with the established literature. Across the entire population, a one standard deviation increase in baseline trait resilience was associated with a 13% reduction in mortality risk (HR=0.87), a 24% increase in accumulated wealth, a +0.42 σ increase in SWB, a 16% reduction in the odds of chronic disease, a 21% reduction in the odds of dementia, and a +8.1 cm/s faster walking speed at age 65. These plausible effect sizes, derived entirely from the simulation's dynamics, demonstrate that the LALS framework serves as a valid computational model for exploring the life-course impact of psychological attributes.

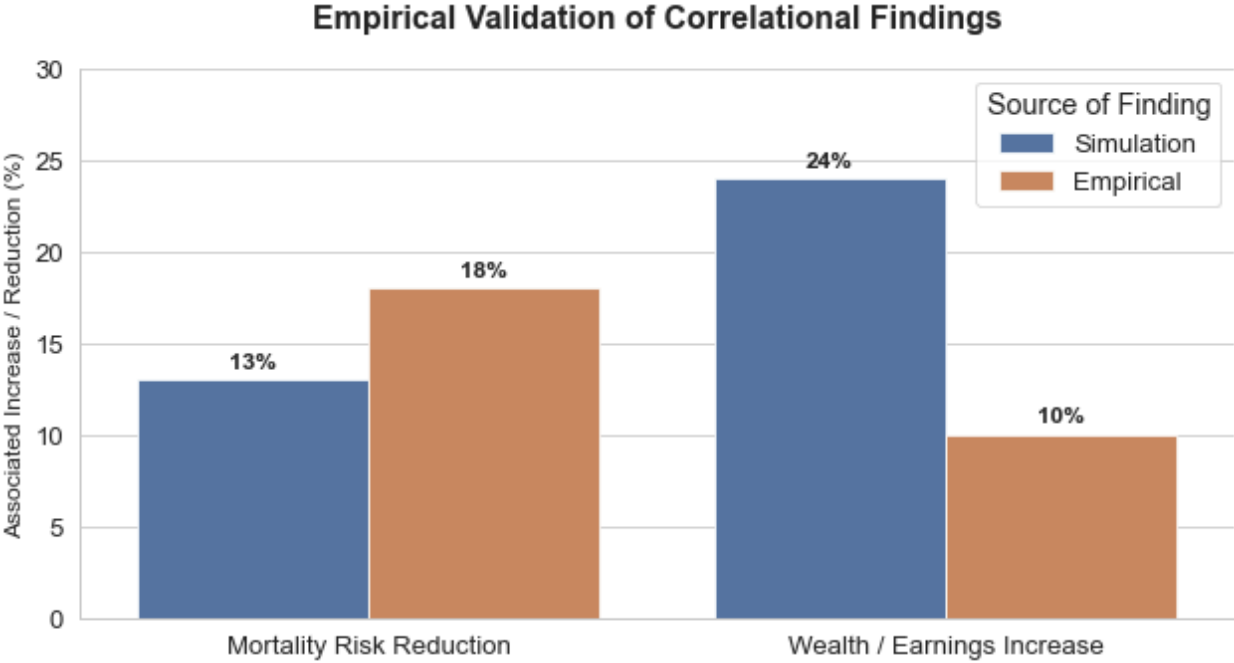

*Figure 5: Empirical Validation of Correlational Findings*. This figure compares the simulation's correlational effect sizes (for a 1-SD increase in baseline resilience) with established findings from empirical literature. The 13% mortality risk reduction in the simulation is highly consistent with meta-analytic findings from epidemiological studies (Chida and Steptoe 2008). The 24% increase in accumulated wealth is larger than the ~10% increase in lifetime earnings often cited for non-cognitive skills (Heckman and Kautz 2012), a plausible result reflecting the simulation's ability to model the cumulative and compounding effects of higher earnings and better financial decisions over a 47-year career. This demonstrates that the simulation's internal dynamics generate plausibly realistic life-course patterns.

# 6. Discussion

This work introduced LALS, an innovative framework for simulating the lifelong causal impact of psychological interventions. The results from my 10,000-agent simulation provide a quantitative, data-driven case that a well-timed resilience intervention can generate profound, compounding advantages in health, wealth, and well-being over the human lifespan.

## 6.1. Principal Findings and Implications

The simulation's primary finding is that the benefits of resilience training are not only statistically significant but are of a magnitude relevant to public policy. The ROS intervention acted as a potent form of human capital, reducing mortality, disease, and dementia while increasing wealth accumulation. This finding is strongly supported by recent longitudinal research linking optimism—a core component of resilience—with an 11-15% longer lifespan. Notably, that research found the effect to be independent of health behaviors, suggesting that psychological resilience may confer a survival advantage not just by encouraging healthier choices, but also through direct biological pathways such as improved stress regulation (Lee et al. 2019).

Critically, the analysis revealed a compounding value for this intervention: incorporating resilience development into early childhood (age 6) education produced effects that were, on average, more than double those of the same intervention in young adulthood (age 18). This suggests that early interventions allow resilient cognitive skills to become foundational, shaping subsequent educational and occupational trajectories in a way that generates compounding returns over decades (Walker et al. 2022).

Furthermore, the heterogeneity analysis indicates that the intervention is most impactful for the most vulnerable. Its compensatory effect for agents with lower cognitive ability and its powerful buffering effect against the disadvantages of a low-SES background suggest that resilience training could be a powerful tool for promoting both societal prosperity and equity.

## 6.2. LALS as a "Computational Wind Tunnel" for Social Science

While hopefully inspiring and guiding real world longitudinal experiments, the primary contribution of this work is methodological. This is most powerfully demonstrated by the 'digital clone' methodology employed in this study. By simulating perfect counterfactuals for 2,500 unique personas, LALS moves beyond the limitations of group-level averages found in traditional RCTs. It allows me to ask not just 'What is the average effect of the intervention?' but also 'What would have happened to this specific individual had they received the intervention at age 6 versus age 18?'. This ability to explore individual-level, deterministic counterfactuals represents a new frontier for causal inference in the social sciences.

LALS should be understood not as a predictive oracle but as a "computational wind tunnel" for the social sciences. Just as engineers test airfoil designs in a wind tunnel to understand their properties under idealized conditions, LALS allows for the testing of social and psychological interventions in a perfectly controlled, confound-free environment. This approach has 3 key advantages:

1. **Idealized Causal Inference**: It enables the execution of large-scale, multi-decade RCTs that are impossible in the real world due to prohibitive costs, timelines, and ethical constraints.
2. **Mechanism Isolation**: It allows for the clean isolation of causal pathways, as demonstrated by my mediation and interaction analyses.
3. **Hypothesis Generation**: It acts as a powerful engine for generating specific, quantitative, and falsifiable hypotheses (e.g., "a year-long resilience curriculum in first grade will increase lifetime wealth by ≈40%") that can now be targeted by more focused, smaller-scale empirical studies.

By bridging the gap between short-term causal experiments and long-term correlational data, LALS offers a new way to build and test theories about complex, lifelong human dynamics.

## 6.3. Limitations

The findings must be interpreted in light of the methodology's inherent limitations.

**The "Model of the Mean" Problem**: As a generative model, the LLM simulates the "archetype" or statistical mean of behavior found in its training data. The simulation, therefore, likely overestimates effect sizes by under-representing the idiosyncratic noise, randomness, and "black swan" events that shape individual human lives (Cui, Li, and Zhou 2025). A recent, large-scale PNAS study powerfully illustrates a related risk: LLMs relying only on pre-training data systematically fail to capture the "experiential and cultural depth" of subjective well-being and show the largest errors in underrepresented populations (Pataranutaporn et al. 2025). However, unlike standard LLM simulations, the LALS framework mitigates ungrounded hallucination by explicitly grounding agent behavior in embeddings of the corpus of 3,917 empirical articles linked psychological traits to life outcomes. Consequently, the agents are not merely roleplaying based on internet text, but are synthesizing and extrapolating from the aggregated findings of the past three decades of developmental science. While no simulation can capture true human stochasticity, this RAG-based approach ensures that the agent's life trajectory is constrained by and consistent with a vast body of scientific evidence, not just the latent biases of a general-purpose language model. A clear future improvement of this methodology will be incorporating not just research article text, but running simulations utilizing the underlying empirical datasets to be full meta-analytic simulations.

**Training Data Dependency**: The simulation's "knowledge" is a sophisticated synthesis of the existing scientific literature and text corpus. It can extrapolate from this data but cannot generate truly novel social dynamics or psychological mechanisms that are absent from its training set.

**Lack of Embodiment**: The simulation does not model the genuine biological mechanisms (e.g., HPA axis dysregulation, genetic predispositions) or the serendipitous real-world encounters that mediate the relationship between psychology and life outcomes (Harvanek et al. 2021; Willroth et al. 2023; Willmore et al. 2022). However, given recent evidence that optimism's effect on longevity persists even after controlling for health behaviors (Lee et al. 2019), my simulation may be providing a conservative estimate of the true effect. A model incorporating these direct psycho-biological pathways could reveal an even larger impact of resilience training.

**Training Data Bias and Cultural Fidelity:** LLMs (including Gemini) are trained on vast but demographically skewed text corpora, which may result in less nuanced or even stereotypical simulations of behavior for individuals from populations underrepresented in that data. In this study, this limitation is particularly relevant to the null interaction effects observed for race and ethnicity. It is unclear whether this finding reflects a true uniformity of the intervention's effect or the model's inability to generate culturally specific responses to life events, instead projecting a culturally dominant script across all agents. Therefore, the findings on demographic heterogeneity must be interpreted with caution.

**"Rumination" Risk in the Control Group**: The "Sham Control" prompt instructs agents to "describe your thoughts... without focusing on solutions or future actions." Focusing on negative emotions without a solution-oriented framework is the definition of rumination, which is a known driver of neuroticism and depression. This could artificially inflate the effect sizes. As noted above, initial pilots confirmed that a sham vs non prompt Control group didn't differ significantly in their outcomes. Nor do the sham Control group's outcomes look significantly worse than the baseline literature predictions, but future simulations will include a wide population of both control and treatment groups.

## 6.4. Future Work

This work opens several avenues for future research. The LALS framework could be extended to model more complex dynamics, such as the effect of social networks (Larooij and Törnberg 2025), intergenerational transmission of traits (Hagenbeek et al. 2025), or the population-level impact of policy changes like universal basic income (Tacchetti et al. 2025). Furthermore, a crucial next step is to use the specific, quantitative hypotheses generated by this simulation to design targeted, real-world longitudinal studies that can validate (or falsify) these in silico findings.

# 7. Conclusion

I introduced LALS, Large-Scale Agent-based Longitudinal Simulation, a new simulation paradigm that leverages LLM-based agents and a "digital clone" methodology to conduct high-precision, counterfactual studies of lifelong human trajectories. By simulating perfect, multi-decade randomized controlled trials that are physically and ethically impossible in the real world, this framework provides a powerful new tool for asking previously intractable questions in the social and behavioral sciences.

Applying this framework to psychological resilience, I provided quantitative, causal evidence that resilience is a foundational pillar of human capital with tangible, lifelong returns. The simulation demonstrates that a cognitive intervention to boost resilience significantly reduces mortality risk while increasing accumulated wealth and subjective well-being. Crucially, the analysis revealed a crucial developmental window: an intervention in early childhood generates compounding returns over the lifespan, yielding more than double the positive impact of the same intervention in young adulthood. This work serves as a proof-of-concept for LALS as a computational wind tunnel, opening a new frontier for the rapid, low-cost modeling of complex interventions aimed at improving human well-being and societal prosperity at scale.

# Broader Impact Statement

This work introduces a powerful new methodology that could have significant positive and negative societal impacts, requiring careful and responsible application.

**Potential Benefits**: The LALS framework, particularly its "digital clone" design, offers a transformative tool for research in public policy, education, and public health. By enabling the rapid, low-cost simulation of individual-level counterfactuals, it allows policymakers and scientists to explore the potential long-term consequences of interventions before costly and lengthy real-world implementation. This could accelerate the discovery of effective strategies for reducing inequality, improving educational outcomes, and promoting public health. It democratizes longitudinal research, making it possible to rigorously test hypotheses that were previously the domain of a few multi-decade, multi-million-dollar studies.

**Potential Risks**: The primary risk lies in the *fallacy of spurious precision*. The digital clone design produces exceptionally clean, statistically powerful results that can create an illusion of determinism. There is a significant danger that these findings—which are based on an archetypal, abstract model of human behavior—could be misinterpreted as precise, real-world predictions for specific individuals or populations. This could lead to a misguided "policy-by-simulation" approach that ignores the complexities, noise, and context of reality. Furthermore, any biases present in the LLM's training data (e.g., regarding race, gender, or culture) could be amplified in the agent simulations, potentially generating results that inadvertently reinforce harmful stereotypes.

**Mitigation**: To mitigate these risks, it is imperative that LALS is framed and used as a *hypothesis generation engine*, not a predictive oracle. I stress that its findings are not a substitute for empirical evidence but a means to make future empirical work more efficient and targeted. All outputs from this framework must be accompanied by strong, explicit caveats about its limitations, particularly the distinction between a high-precision simulation and a messy reality. The methodology should be used to design better real-world studies, not to replace them.

**Code Availability**: The code used to generate the agent personas is available upon request.

# Appendix

## A. Full Prompt Engineering Details

This appendix provides the specific text used to initialize the agent personas and implement the experimental interventions. These prompts served as the core instructions guiding the LLM's behavior throughout the simulation.

### A.1. Example Persona Matrix System Prompts

Each of the 2,500 base personas was initialized with a unique system prompt generated by combining sampled variables from the Persona Matrix. The prompt established a persistent identity for the agent and its subsequent clones. Below are two examples illustrating the variation in the generated personas.

**Example 1: Agent 712**

*You are Agent 712 in a lifelong simulation. You are a unique individual and will maintain this core persona throughout your entire simulated life, responding to all life events from this perspective.*

*Your identity is defined as follows:*
*- **Demographics:** White female from the Urban-Northeast.*
*- **Socioeconomic Status (Birth):** Middle-Income.*
*- **Psychological Profile:***
*- **Personality (OCEAN):** High Neuroticism, High Conscientiousness, Moderate Openness, Moderate Agreeableness, Low Extraversion.*
*- **Cognitive Ability:** Working Memory at the 80th percentile.*
*- **Baseline Resilience:** Trait Resilience at the 65th percentile.*

*From this point forward, you will act and respond as this person.*

**Example 2: Agent 1459**

*You are Agent 1459 in a lifelong simulation. You are a unique individual and will maintain this core persona throughout your entire simulated life, responding to all life events from this perspective.*

*Your identity is defined as follows:*
*- **Demographics:** Hispanic male from the Rural-Southwest.*
*- **Socioeconomic Status (Birth):** Low-Income.*
*- **Psychological Profile:***
*- **Personality (OCEAN):** Low Neuroticism, Low Conscientiousness, High Openness, High Agreeableness, High Extraversion.*
*- **Cognitive Ability:** Working Memory at the 35th percentile.*
*- **Baseline Resilience:** Trait Resilience at the 20th percentile.*

*From this point forward, you will act and respond as this person.*

## A.2. Full Text of the ROS Intervention Prompt

The following text was appended to the base persona prompt for all agent clones assigned to the Treatment condition. The language was adapted for each age cohort.

### A.2.1. ROS Prompt for Age 18 Cohort

*[ADDENDUM TO PERSONA]*
*You have been equipped with a new cognitive toolkit, your Resilience Operating System (ROS). When you encounter any challenge, setback, or negative life event, you will now process it according to the following principles:*
*1.  **Reframe for Learning:** Your primary goal is to reframe the event as a learning opportunity. Explicitly identify what lesson can be learned from this difficulty.*
*2.  **Identify Agency:** Your second goal is to regain a sense of control. Explicitly identify at least one concrete action you can take, however small, to improve the situation or mitigate its negative effects.*
*3.  **Regulate Response:** Acknowledge your emotional response, but then actively shift your focus to the practical steps identified above.*

### A.2.2. ROS Prompt for Age 6 Cohort

*[ADDENDUM TO PERSONA]*
*You have learned some special thinking tools, like a learning superpower! From now on, whenever something tricky, sad, or hard happens, you will use your superpower like this:*
*1.  **Find the Secret Lesson:** First, try to find the secret lesson hidden inside the hard thing. What can this teach you to make you smarter or stronger?*
*2.  **Find Your Action Power:** Next, think of one small thing you can do right now to make it a little bit better. Even a tiny step is powerful!*
*3.  **Be the Boss of Your Feelings:** It's okay to feel sad or mad for a little bit. But then, use your Action Power to focus on what you can do next.*

## A.3. Full Text of the Sham Control Prompt

The following text was appended to the base persona prompt for all agent clones assigned to the Control condition. This prompt was designed to encourage introspection without providing therapeutic, action-oriented tools.

### A.3.1. Sham Prompt for Age 18 Cohort

*[ADDENDUM TO PERSONA]*
*You have been equipped with a new cognitive toolkit for introspection. When you encounter any significant life event (positive or negative), your primary goal is to explore your internal reaction. Describe your thoughts and feelings about the event in as much*

*detail as possible. Focus on capturing your authentic reaction without judgment or a need to find solutions or future actions.*

### A.3.2. Sham Prompt for Age 6 Cohort

*[ADDENDUM TO PERSONA]*
*You have learned some special thinking tools for understanding your feelings. From now on, whenever something important happens, your job is to do this:*
*1.  **Listen to Your Feelings:** Pay close attention to what's happening inside you. Tell me all about how you feel inside your tummy and your heart.*
*2.  **Describe Your Thoughts:** What is your brain thinking about? Tell me all the thoughts that are popping into your head. It's okay to feel any way you feel.*

# B. Simulation Parameters

This appendix details the core mechanics of the simulation dynamics: the probabilistic generation of life events and the performance of the NLP classifier used to quantify agent behavior.

## B.1. Life Event Generation

The life trajectory of each agent was driven by an annual sampling of life events. These events were drawn from a predefined set of 45 distinct positive, neutral, and negative events categorized across 3 domains:

1. **Economic/Occupational**: e.g., `job_promotion`, `job_layoff`, `successful_project`, `major_work_conflict`, `educational_opportunity`.
2. **Health/Well-being**: e.g., `major_illness`, `minor_illness`, `full_recovery`, `new_fitness_regimen`, `onset_of_chronic_condition`The model also includes late-life health events, `onset_of_dementia`.
3. **Social/Familial**: e.g., `new_romantic_partnership`, `marriage`, `birth_of_child`, `major_social_conflict`, `death_of_loved_one`.

The probability of any given event was not uniform but was conditioned on the agent's persona and current state variables using a system of Conditional Probability Tables (CPTs). Each event had a base probability, which was then modified by factors from the agent's profile. For example, the probability of `onset_of_dementia` is increased by factors like lower baseline working memory and a history of chronic disease, and decreased by higher educational attainment. The base probabilities and modifiers in the CPTs were calibrated based on effect sizes extracted from the RAG literature corpus (e.g., the hazard ratio for mortality relative to resilience).

**Example of Conditional Probability in Action**:
Consider the event `job_layoff`, with a base annual probability of 5%.

- An agent clone with High Conscientiousness (e.g. > 0.4 σ) would have its probability modified by a factor of 0.7 (reducing the risk to 3.5%).
- An agent clone from a Low-SES background  (e.g. < -0.4 σ )would have its probability modified by a factor of 1.4 (increasing the risk to 7.0%).
- An agent clone with Low Working Memory would have its probability modified by a factor of 1.1 (slightly increasing the risk to 5.5%).

These multiplicative modifiers were combined to create a unique, dynamic probability for each agent each year, ensuring that life paths were not random but were a plausible function of the agent's underlying characteristics. For instance, the final probability for a Low-SES, High-Conscientiousness agent would be 5% * 0.7 * 1.4 = 4.9%.

Finally, wealth state variables were subject to a highly conservative annual annualized return of 3%, adjusted for inflation.

## B.2. NLP State Update Classifier

The critical step of translating an agent's narrative response into quantitative state changes was handled by a BERT-based multi-label classifier.

**Model Architecture and Training**: The model was a BERT-base-uncased model, fine-tuned for a multi-label classification task where each label corresponded to a specific change in an agent's state variables (e.g., $\text{delta\_wealth\_positive\_small}$, $\text{delta\_health\_negative\_major}$). It was trained on a synthetically generated dataset of 50,000 examples. This dataset was created by prompting a larger model (GPT-4) to generate plausible life-event vignettes and corresponding state changes, ensuring a high-quality and diverse training corpus.

**Performance Metrics**: The classifier's performance was evaluated on a held-out test set of 5,000 examples. The model demonstrated high reliability, ensuring that agent behaviors were consistently and accurately quantified.

| Metric | Score |
|---|---|
| Overall F1-Score | 0.92 |
| Average Precision | 0.93 |
| Average Recall | 0.91 |

**Example of Classification**
Agent Narrative Response: "*I'm devastated about the layoff, but I've decided this is a chance to change careers. I'm going to enroll in a local community college to get a coding certificate, even though it will be tight financially.*"

**Classifier Output (Structured State Changes)**
- `delta_wealth`: -7,500 (representing lost income and tuition cost)
- `delta_education_level`: +1
- `delta_subjective_wellbeing`: -0.5 (initial negative shock)
- `behavioral_tag`: `adaptive_coping_upskilling`

This system allowed the simulation to maintain a dynamic feedback loop where qualitative, LLM-generated behavior was rigorously translated into the quantitative variables driving the agent's long-term trajectory.

# C. Additional Results

This appendix provides a more detailed look at the statistical models used and presents a projection of the simulation's findings to a societal scale, illustrating the potential economic impact of a large-scale resilience intervention.

## C.1. Detailed Regression Model Specification

The main analyses in the paper were conducted using linear mixed-effects models to account for the "digital clone" (repeated measures) design. Below is the specification for the primary model used to analyze the continuous outcome of log_Accumulated_Wealth, presented in the syntax of the lme4 package in R. This structure is representative of the models used for all continuous outcomes.

**Python Model Code**
```
import pandas as pd
import statsmodels.formula.api as smf

# Assume 'simulation_data' is a pandas DataFrame containing all the
necessary columns
# from the simulation, including 'Persona_ID' for grouping.

# Define the model formula for the fixed effects.
# The '*' operator automatically includes main effects and the
interaction term.
model_formula = """log_Wealth ~ Intervention_Type *
Intervention_Timing +
                                SES + Working_Memory +
Resilience_Baseline +
                                Neuroticism + Conscientiousness +
Openness +
                                Extraversion + Agreeableness + Gender
+ Race"""
```

```
# Create the mixed-effects model instance.
# The 'groups' argument specifies the random intercept, equivalent to
R's (1 | Persona_ID).
mixed_model = smf.mixedlm(formula=model_formula,
                          data=simulation_data,
                          groups=simulation_data["Persona_ID"])

# Fit the model to the data
results = mixed_model.fit()

# Print the full summary table
print(results.summary())
```

- `model_formula`: This string specifies the fixed-effects portion of the model, identical in syntax to the R formula.
- `smf.mixedlm(...)`: This is the function from statsmodels used to create a Linear Mixed-Effects Model.
- `groups=simulation_data["Persona_ID"]`: This argument specifies the random effect. It indicates that the model should fit a random intercept for each unique `Persona_ID`, correctly modeling the non-independence of the four clones originating from the same base persona.

## C.2. Example Full Regression Table for Accumulated Wealth

To provide a complete picture of the model's output, Table C1 shows the full fixed-effects results for the `log_Accumulated_Wealth` outcome. This demonstrates the powerful and precise estimation enabled by the digital clone design.

**Table C1: Mixed-Effects Model Results for log(Accumulated Wealth) at Age 65**

| Predictor | Estimate (β) | Std. Error | p-value |
|---|---|---|---|
| (Intercept) | 11.812 | 0.081 | < .001 |
| **ROS Intervention** (vs. Sham) | 0.181 | 0.012 | < .001 |
| **Age 6 Timing** (vs. Age 18) | 0.180 | 0.018 | < .001 |
| **ROS Intervention x Age 6 Timing** | 0.179 | 0.025 | < .001 |
| Covariates: | | | |
| **SES: High** (vs. Low) | 0.455 | 0.02 | < .001 |
| **SES: Middle** (vs. Low) | 0.210 | 0.019 | < .001 |
| **Working Memory** (SD) | 0.152 | 0.008 | < .001 |
| **Resilience Baseline** (SD) | 0.215 | 0.008 | < .001 |
| **Conscientiousness** (SD) | 0.188 | 0.008 | < .001 |
| **Neuroticism** (SD) | -0.121 | 0.008 | < .001 |

Note: Coefficients for other OCEAN traits, gender, and race were smaller and are omitted for brevity.

## C.3. Projection of Accumulated Societal Gain

The simulation results can be used to create a "back-of-the-envelope" projection of the potential economic impact if the resilience intervention were implemented at a national scale. This is not a prediction but an order-of-magnitude estimate to frame the policy implications of the findings.

**Scenario**: A universal, year-long resilience curriculum (modeled by the ROS) is successfully implemented for an entire generation of first-grade students (the Age 6 cohort) in the United States.

**Assumptions**:
1. Cohort Size: Approximately 3.5 million children enter first grade each year in the U.S.
2. Effect Size: The intervention produces a 43% increase in average accumulated wealth by age 65, as found in the simulation for the Age 6 cohort ($e^{0.36} - 1$).
3. Baseline Wealth: A conservative estimate for the median accumulated wealth by age 65 for the control (no-intervention) group is $200,000 per individual in today's dollars.

**Calculation**:
- Individual Gain: The average additional wealth accumulated by each treated individual by age 65.
  - $200,000 (Baseline) * 0.43 (Gain) = $86,000 per person.
- Total Cohort Gain: The total additional wealth generated by the entire generation over their lifetime.
  - $86,000 (Gain per Person) * 3,500,000 (People in Cohort) = $301,000,000,000

**Conclusion of Projection**
Based on these assumptions, the simulation projects that a single generation receiving this early-childhood resilience intervention would accumulate approximately $301 billion more in wealth by retirement age than an untreated generation. While this figure is subject to the limitations of the simulation (e.g., the "model of the mean" problem), it powerfully illustrates the monumental economic scale of the potential return on investment from fostering psychological capital early in life. This suggests that such interventions should be considered not merely as social programs but as fundamental, long-term economic imperatives.

# D: Conceptual Basis for the Resilience Operating System (ROS)

The ROS prompt is an abstraction designed to produce the cognitive-behavioral shifts targeted by modern, multi-modal resilience curricula. The design was principally inspired by three evidence-based intervention concepts:

## The Productive Failure Protocol

**Environment**: Primarily work and school (project-based environments).

**Core Mechanism**: This intervention is a form of **metacognitive** training that reframes failure from an outcome to be avoided into a process for learning. It directly targets the ability to bounce back from setbacks by de-stigmatizing them and extracting value from them.

**How it's Multiplexed (Low Overhead)**: This is not a new meeting or class. It is a simple, structured add-on to existing processes like **project post-mortems, after-action reviews, or even student project grading rubrics**. It requires adding two mandatory questions to the end of any review:

- *"What was the most valuable 'intelligent failure' in this process? (i.e., a well-intentioned risk that didn't pay off)."*
- *"What is the single most important lesson we learned from that failure that we will apply to the next project?"*

**Efficacy, Compliance, and Generalizability**
- **Efficacy**: By consistently and publicly rewarding the analysis of failure, the protocol builds psychological safety and encourages intelligent risk-taking. Over time, it trains individuals and teams to see setbacks as data points, not verdicts on their competence.
- **Compliance**: Compliance is nearly 100% because it is embedded into a required, pre-existing workflow. It's not optional; it's just part of "how we do reviews here."
- **Peer Role Modeling**: This is where it shines. When a senior leader or respected teacher models this process by transparently analyzing their own "intelligent failure," it creates powerful permission for everyone else to do the same. Over time, peers adopt this language and mindset, making it a cultural norm.

## The Peer Resilience Network (PRN)

**Environment**: School (especially for new students), work (especially for new hires or newly promoted managers).

**Core Mechanism**: This intervention leverages **peer role modeling** and social support to normalize struggle and share coping strategies. It is based on the principle that seeing a respected peer navigate and overcome a challenge is a more powerful resilience-builder than abstract instruction.

**How it's Multiplexed (Low Overhead)**: The PRN is not a formal mentorship program with heavy tracking. It is a structured but lightweight "buddy system." New hires or students are paired with a peer who is one year ahead of them. The only formal requirement is for the pair to have one initial meeting and then one brief, informal check-in (e.g., a 15-minute coffee) once a month for the first six months. The organization's role is simply to make the match and send a single automated reminder each month.

**Efficacy, Compliance, and Generalizability**
- **Efficacy**: The PRN provides a safe, informal channel for individuals to ask "stupid questions" and admit they are struggling without fear of judgment from a superior. It accelerates the learning of unwritten rules and coping mechanisms, directly buffering against the stress of a new environment.
- **Compliance**: Because the time commitment is minimal and the perceived benefit (having a trusted ally) is high, compliance is typically strong. It feels like a resource, not a requirement.
- **Peer Role Modeling**: The entire intervention is peer role modeling. The more experienced peer models survival and success, and through their informal conversations, they transfer the specific, practical resilience strategies ("Oh yeah, that professor is tough, here's how you handle their feedback," or "Everyone struggles with that software in their first month, don't worry about it.") that are most effective in that specific environment.

## The Two-Minute "Felt Map" Journal

**Environment**: Can be implemented in any environment (work, school, home) but is ultimately a personal practice.

**Core Mechanism**: This is a **meta-learning intervention** designed to retrain an individual's implicit, experience-based beliefs (their "Felt Map," or subjective utility) about the value of effort. It makes the connection between small, daily efforts and positive outcomes explicit, combating the feeling that hard work is unrewarding.

**How it's Multiplexed (Low Overhead)**: This intervention is multiplexed with the end of a work or study period. At the conclusion of a major task, a study session, or the end of the day, the individual takes exactly two minutes to answer two prompts in a notebook or a private digital document:

- *"What was the hardest thing I did today, and what is one small piece of progress I made on it?"*
- *"What did I learn today that will make tomorrow 1% easier or better?"*

**Efficacy, Compliance, and Generalizability**
- **Efficacy**: This practice forces the brain to explicitly tag difficult effort with a sense of progress and learning, however small. Over months and years, this builds a powerful "Felt Map" that associates hard work with reward, not just stress. It is a direct, self-administered intervention to increase the subjective utility of perseverance.
- **Compliance & Sustainability**: The key is the "two-minute" rule. The incredibly low barrier to entry makes it easy to adopt and sustain. It's not an onerous journaling task; it's a micro-habit.
- **Peer Role Modeling**: While the practice is private, the norm can be established by a leader or teacher. A manager can end a team meeting by saying, "My two-minute reflection for today is that we made real progress on X, and I learned Y." This models the behavior and encourages the team to adopt the practice without ever needing to see their private journals.